\documentclass[%
 reprint,
 amsmath,amssymb,
 aps,prb
]{revtex4-1}

\usepackage{graphicx}
\usepackage{dcolumn}
\usepackage{bm}
\usepackage[bookmarks=true,colorlinks,linkcolor=blue,urlcolor=blue,citecolor=blue]{hyperref}

\renewcommand{\vec}[1]{\mathbf{#1}} 

\begin{document}

\title{Quantum spin liquid ground states of the Heisenberg-Kitaev model on the triangular lattice }

\author{Pavel Kos}
\affiliation{Physics Department, Arnold Sommerfeld Center for Theoretical Physics, and Center for NanoScience, Ludwig-Maximilians University Munich, Germany}
\author{Matthias Punk}
\affiliation{Physics Department, Arnold Sommerfeld Center for Theoretical Physics, and Center for NanoScience, Ludwig-Maximilians University Munich, Germany}

\date{\today}

\begin{abstract}

We study quantum disordered ground states of the two dimensional Heisenberg-Kitaev model on the triangular lattice using a Schwinger boson approach. Our aim is to identify and characterize potential gapped quantum spin liquid phases that are stabilized by anisotropic Kitaev interactions. For antiferromagnetic Heisenberg- and Kitaev couplings and sufficiently small spin $S$ we find three different symmetric $Z_2$ spin liquid phases, separated by two continuous quantum phase transitions. Interestingly, the gap of elementary excitations remains finite throughout the transitions. The first spin liquid phase corresponds to the well known zero-flux state in the Heisenberg limit, which is stable with respect to small Kitaev couplings and develops $120^\circ$ order in the semi-classical limit at large $S$. In the opposite Kitaev limit we find a different spin liquid ground-state, which is a quantum disordered version of a magnetically ordered state with antiferromagnetic chains, in accordance with results in the classical limit. Finally, at intermediate couplings we find a spin liquid state with unconventional spin correlations. Upon spinon condensation this state develops Bragg peaks at incommensurate momenta in close analogy to the magnetically ordered Z2 vortex crystal phase, which has been analyzed in recent theoretical works. 
\end{abstract}

\maketitle

\section{Introduction}

In recent years it has been realized that spin-orbit coupling can drive a wealth of interesting new phenomena in electronic systems. Prominent examples are topological insulators\cite{RevModPhys.82.3045, RevModPhys.83.1057} and Weyl semi-metals,\cite{PhysRevB.83.205101} which exhibit exotic edge states due to their non-trivial topological band structure. Spin-orbit coupling can also play an important role in strongly correlated electron systems.\cite{Witczak, Rau} In particular the Mott insulating iridium oxides have been at the focus of experimental and theoretical research. Here, the crystal field splitting of partially filled $5d$ orbitals of Ir$^{4+}$ ions together with strong spin-orbit coupling gives rise to half-filled $j=1/2$ Kramers doublets, which can form a Mott insulator with an effective spin-1/2 degree of freedom.\cite{PhysRevLett.101.076402} Interestingly, the exchange coupling between these local moments can be highly anisotropic and depends on  the spatial direction of the exchange path, leading to Kitaev-type interactions in such materials.\cite{PhysRevLett.102.017205, PhysRevLett.105.027204}
From a theoretical perspective models with Kitaev-type exchange couplings are particularly interesting, because they can host exotic quantum spin liquid ground states. \cite{Kitaev20062} 

While most iridate compounds that have been studied so far are honeycomb lattice based,\cite{PhysRevLett.102.256403, PhysRevB.82.064412, PhysRevLett.108.127203, PhysRevLett.108.127204} Ba$_3$IrTi$_2$O$_9$ is a layered triangular lattice material and thermodynamic measurements indicate that the ground state is potentially a quantum spin liquid.\cite{PhysRevB.86.140405} Indeed, susceptibility and specific heat measurements show no sign of magnetic ordering down to temperatures of $0.35 K$, even though the exchange coupling is rather large, as evidenced by a large Curie-Weiss temperature of $\theta_\text{CW} \simeq -130 K$. Recent theoretical works have argued that the low energy properties of the $j=1/2$ moments can be modelled by a Heisenberg-Kitaev Hamiltonian on the triangular lattice,\cite{PhysRevB.91.155135} potentially with additional off-diagonal exchange terms.\cite{PhysRevB.92.165108}
The interplay of Heisenberg- and Kitaev interactions on the triangular lattice can stabilize interesting magnetically ordered states. In particular, for antiferromagnetic couplings a Z2 vortex crystal state with non-coplanar spin correlations has been shown to occupy a large part of the phase diagram in the classical limit,\cite{PhysRevB.93.104417} as well as in the quantum model.\cite{PhysRevB.91.155135, Shinjo} This vortex crystal can be understood as a state with $120^\circ$ order, whose topological vortex excitations are condensed and form triangular lattice crystal. Other numerical work suggests that a chiral spin liquid ground state may be realized in the vicinity of the Kitaev point.\cite{1367-2630-17-4-043032}

Motivated by these experimental and theoretical findings we study potential quantum disordered spin liquid phases of the Heisenberg-Kitaev model on the triangular lattice in the following. Using Schwinger-Boson mean-field theory (SBMFT) we construct gapped spin liquid states and compute spin-correlation functions to characterize their properties. We find three distinct symmetric $Z_2$ spin liquids which realize different symmetry enriched topological (SET) phases,\cite{PhysRevB.87.104406, PhysRevB.87.155115, PhysRevB.87.165107, PhysRevB.93.155121} but are all characterized by the same projective symmetry group (PSG).\cite{PhysRevB.65.165113} Interestingly, these states are separated by continuous quantum phase transitions where the gaps to all elementary excitations remain finite at the transition. Continuous transitions between different topological states without closing the gap have been discussed in the context of topological insulators,\cite{Ezawa:2013aa} but so far have not been considered for SET phases with bulk topological order.  

It is worth noting that even though SBMFT doesn't give quantitatively accurate results, it is a very useful tool to construct and characterize potential spin liquid states in Mott insulators and to analyze their qualitative behavior.\cite{PhysRevB.45.12377,PhysRevB.74.174423} Furthermore, it allows us to make connections to earlier work on magnetically ordered states by studying the semi-classical large spin $S$ limit.

The outline of this article is as follows: in Sec.~\ref{sec2} we introduce the Heisenberg-Kitaev model and develop a decoupling of the Kitaev interaction in terms of triplet bond operators within the SBMFT approach. Furthermore we perform a PSG analysis to identify suitable mean-field ans\"atze. In the remainder of this work we focus on the only fully symmetric, time-reversal invariant ansatz and our results are presented in Sec.~\ref{sec3}, where we determine the phase diagram for antiferromagnetic Heisenberg- and Kitaev couplings and compute spin structure factors to characterize the three different spin liquid phases that we found. In addition, we show that our approach correctly reproduces the previously studied magnetically ordered phases in the Kitaev- and Heisenberg limits for large spins $S$ and comment on the nature of the phase transitions between the different $Z_2$ spin liquids. We conclude with a discussion in Sec.~\ref{sec4}. A brief discussion of generalized, weakly symmetric PSG ans\"atze where time-reversal and parity symmetries are broken can be found in Appendix.~\ref{appendix:WSansatz}.

\section{MODEL AND METHODS}
\label{sec2}

\subsection{The Heisenberg-Kitaev Model on the Triangular Lattice}
\begin{figure}
\centering
\includegraphics[width=0.6\columnwidth]{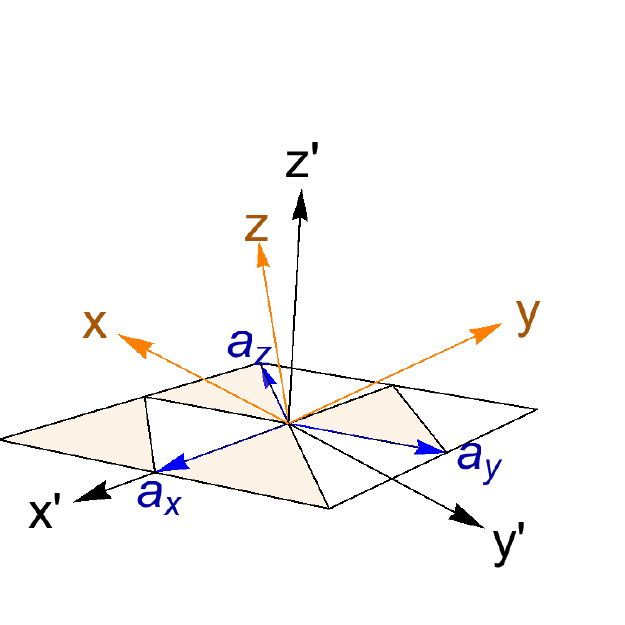}
\caption{(Color online) Basis vectors $\vec{a_i}$ of the triangular lattice and definition of the coordinate system for the spin degrees of freedom.}
\label{fig:triangular}
\end{figure}

The Heisenberg-Kitaev model on the triangular lattice is given by the following Hamiltonian
\begin{align}
H_{\text{HK}}&= J_H \sum_{\langle ij \rangle} \vec{S_i} \cdot \vec{S_j} + J_K \sum_{\gamma||\langle ij \rangle } S_i^\gamma S_j^\gamma,
\label{HHK}
\end{align}
where $\vec{S_i}$ is a spin-1/2 operator located on lattice site i and the sums run over nearest neighbor sites. The first term describes the usual isotropic Heisenberg interaction, whereas the second, Kitaev-type interaction term explicitly breaks spin-rotation invariance. It couples only the $\gamma \in \left\{x,y,z\right\}$ component $S^\gamma$ of the spin operators on bonds with direction $\vec{a}_\gamma$ (see Fig.~\ref{fig:triangular}).
In the following we restrict our discussion to antiferromagnetic couplings $J_H,J_K>0$ and parametrize the interactions using an angular variable
\begin{align} J_H=J \cos{\psi},&&J_K= J \sin{\psi} \ .
\end{align} 
The units of energy will be set by $J=\sqrt{J_H^2+J_K^2}=1$. Using the Klein duality our results also describe a specific parameter regime with ferromagnetic Heisenberg couplings.\cite{PhysRevB.89.014414}

The triangular lattice is spanned by the basis vectors $\vec{a_x}= \vec{e_x'}$, $\vec{a_y}= - \frac{1}{2}\vec{e_x'}+\frac{\sqrt{3}}{2}\vec{e_y'} $, where the lattice constant has been set to unity. Additionally we define $\vec{a_z}= - \frac{1}{2}\vec{e_x'}-\frac{\sqrt{3}}{2}\vec{e_y'} =-\vec{a_x}-\vec{a_y}$. Here we expressed the vectors in the primed coordinate system, where $\vec{e_x'}$ and $\vec{e_y'}$ is a pair of orthogonal vectors in the lattice plane and $\vec{e_z'}$ is perpendicular to the lattice.

Spin-orbit coupling locks the primed coordinate system to the unprimed one, which defines the orientation of the spin operators with respect to the lattice, i.e.~the component $S^\gamma$ points in the $\gamma$ direction of the unprimed coordinate system, as shown in Fig.~\ref{fig:triangular}. This coordinate system is fixed by the condition $\vec{e_z'}=(1,1,1)/\sqrt{3}$ and that $\vec{e_x}$ projected onto the lattice plane points into the direction of $\vec{e_x'}$. 
Note that the combined spin-orbit symmetry is $D_{3d}$,\cite{PhysRevB.93.104417} and threefold rotations $C_3$ around the $\vec{e_z'} \sim (1,1,1)$ axis act as:
\begin{subequations}
\begin{align}
&C_3: (\vec{a_x},\vec{a_y},\vec{a_z}) \to (\vec{a_y},\vec{a_z},\vec{a_x}),\\
&C_3: (S^x,S^y,S^z) \to (S^y,S^z,S^x).
\end{align}
\end{subequations}

\subsection{Schwinger Boson Mean-Field Theory (SBMFT)}

Schwinger-Boson Mean Field Theory is a useful analytical approximation to study quantum disordered ground-states of correlated spin models. It is based on representing spin operators with bosons and decoupling the interactions using suitably chosen bond operators. Subsequently, the theory is analyzed using a mean-field approximation by making an ansatz for the expectation values of the bond operators, which are computed self-consistently. This approximation can be formally justified in a large $N$ limit in generalized models with $Sp(N)$ symmetry.\cite{PhysRevLett.66.1773, PhysRevB.45.12377}

We start by expressing the spin operators in terms of Schwinger bosons as 
\begin{align}
\vec{S_i}=\frac{1}{2} b_{i \alpha}^{\dagger} \vec{\sigma_{\alpha \beta}} b_{i \beta},
\end{align}
where the indices $\alpha$ and $\beta$ run over the $2S+1$ angular momentum basis states (we employ a summation convention over Greek indices) and $\sigma_{\alpha \beta}$ is a $2S+1$ dimensional representation of $SU(2)$. This mapping preserves the angular momentum algebra if there are precisely $2S$ bosons per lattice site 
\begin{align}
\hat{n}_i=\sum_\alpha b_{i \alpha}^{\dagger}  b_{i \alpha}=2S.
\label{lagrange}
\end{align}
Unphysical states with more or less than $2S$ bosons per site need to be projected out.
We impose the above constraint by adding a Lagrange multiplier term $ \sum_i \lambda_i \big( b_{i \alpha}^\dagger b_{i \alpha}-2S \big)$ to the Hamiltonian. 

Focusing on the relevant $S=1/2$ case, the common $SU(2)$ symmetric Heisenberg term of $H_\text{HK}$ can be decoupled in terms of the $SU(2)$ invariant bond operators
\begin{align}
\hat{A}_{ij}=\frac{1}{2}\epsilon_{\alpha \beta} b_{i \alpha}b_{j \beta}, &&
\hat{B}_{ij}=\frac{1}{2} b_{i \alpha}^\dagger b_{j \alpha} \ ,
\end{align}
where $\epsilon_{\alpha\beta}$ is the fully antisymmetric tensor. The non-$SU(2)$ invariant Kitaev interactions can be expressed in terms of the triplet operators
\begin{subequations}
\begin{align}
\hat{t}_{ij}^x&=\frac{i}{2} (b_{i \uparrow}b_{j \uparrow}-b_{i \downarrow} b_{j \downarrow}) \\
\hat{t}_{ij}^y&=\frac{-1}{2} (b_{i \uparrow}b_{j \uparrow}+b_{i \downarrow} b_{j \downarrow})\\
\hat{t}_{ij}^y&=\frac{-i}{2} (b_{i \uparrow}b_{j \downarrow}+b_{i \downarrow} b_{j \uparrow}) \ .
\end{align}
\label{def of t}
\end{subequations}
Note that we included factors of $i$ in the definition of the triplet bond operators for convenience, such that they transform under time reversal only by complex conjugation, where the time reversal operator is given by $T=- i \sigma^{y} K$ and $K$ denotes complex conjugation.
The interaction terms in the Hamiltonian $\eqref{HHK}$ can now be written as (: : denotes normal ordering)
\begin{subequations}
\begin{align}
\vec{S_i} \cdot \vec{S_j} &= \ :\hat{B}_{ij}^{\dagger} \hat{B}_{ij}:-\hat{A}_{ij}^\dagger \hat{A}_{ij},\\
S_i^\gamma S_j^\gamma &=
  -\hat{t}_{ij}^{\gamma \dagger} \hat{t}_{ij}^\gamma + :\hat{B}_{ij}^{ \dagger} \hat{B}_{ij}: .
\end{align}
\end{subequations}
A similar decomposition has been used in Ref.~[\onlinecite{PhysRevB.86.205124}] for Heisenberg-Kitaev models on the honeycomb lattice.
We are now ready to preform a mean field decoupling by expanding $H_\text{HK}$ to linear order in fluctuations around the expectation values of the various bond operators, e.g. 
\begin{align}
\hat{A}_{ij}^\dagger \hat{A}_{ij} = \hat{A}_{ij}^\dagger A_{ij}+A_{ij}^* \hat{A}_{ij}-|A_{ij}|^2+\mathcal{O} \left( \delta\hat{A}_{ij}^2 \right) \ ,
\end{align}
where $A_{ij}=\langle \hat{A_{ij}} \rangle$ denotes the expectation value. 
The mean-field decoupled Hamiltonian now reads
\begin{widetext}
\begin{align}
H_\text{MF}&=(J_H+J_K) \sum_{\langle ij \rangle} 
\Big[(B_{ij})^*\frac{1}{2} b_{i \alpha}^\dagger b_{j \alpha}+ B_{ij} \frac{1}{2} b_{i \alpha} b_{j \alpha}^\dagger -|B_{ij}|^2 \Big] 
-J_H \sum_{\langle ij \rangle}
\Big[(A_{ij})^*\frac{1}{2} \epsilon_{\alpha \beta} b_{i \alpha}b_{j \beta}+ A_{ij} \frac{1}{2} \epsilon_{\alpha \beta} b_{j \beta}^\dagger b_{i \alpha}^\dagger  -|A_{ij}|^2 \Big]
\nonumber \\
&-J_K \sum_{\gamma||\langle ij \rangle} 
\Big[(t_{ij}^\gamma)^* \hat{t}_{ij}^\gamma
+ t_{ij}^\gamma  (\hat{t}_{ij}^\gamma)^\dagger
 -|t_{ij}^\gamma|^2 \Big]
+ \sum_i \lambda_i \big( b_{i \alpha}^\dagger b_{i \alpha}-2S \big),
\label{hmf0}
\end{align}
\end{widetext}
with $\hat{t}_{ij}^\gamma$ defined in equation \eqref{def of t}. 
Note that the expectation values of the bond operators need to be determined self-consistently:
\begin{align}
A_{ij}=\langle \hat{A}_{ij} \rangle_\text{MF} && 
B_{ij}=\langle \hat{B}_{ij} \rangle_\text{MF} && 
t_{ij}^\gamma=\langle \hat{t}_{ij}^\gamma \rangle_\text{MF} && 
\end{align}
These conditions are equivalent to demanding that we are at a saddle point of the free energy $F_\text{MF}(A_{ij},B_{ij},t_{ij}^\gamma,\lambda_i)$. In order to reduce the complexity of the problem we furthermore make a standard approximation and set $\lambda_i = \lambda$, i.e.~the constraint of having one boson per lattice site is only satisfied on average.

It is important to emphasize that the SBMFT approach has the advantage that we can formally set the spin $S$ to unphysical values $S<1/2$ by adjusting the Lagrange multiplier term specified below Eq.~\eqref{lagrange}. This allows us to reach parameter regimes where quantum fluctuations are artificially enhanced and any kind of magnetic order is suppressed. Note that in the problem considered here, the ground states are always magnetically ordered at the physical value $S=1/2$ within SBMFT. Nevertheless, it is possible that SBMFT underestimates quantum fluctuations and the true ground state of the model \eqref{HHK} at $S=1/2$ is indeed a spin liquid, which might correspond to a state that appears in SBMFT for $S<1/2$. In this case the strength of SBMFT is to provide an analytic tool which allows us to characterize these potential spin liquid phases.

Without constraining the mean-field bond parameters by symmetries, their number grows linearly with system size, which makes the numerical problem of finding the saddle point values cumbersome. If we expect a homogeneous solution, the parameters should attain just a few different values, however. Indeed, numerical solutions of the Heisenberg model on small clusters confirm this expectation.\cite{PhysRevB.86.245132}

We will simplify our problem by choosing an ansatz for the mean-field parameters that is invariant under the symmetries of the original Hamiltonian $H_\text{HK}$. Since our description in terms of Schwinger bosons contains a $U(1)$ gauge redundancy $b_{j\alpha} \to b_{j\alpha} e^{i \varphi_j}$, symmetries can act on $b, b^\dagger$ projectively, i.e.~the bond parameters should only be invariant under lattice symmetries modulo gauge transformations. All compatible ans\"aze can be determined by the projective symmetry group approach (PSG).\cite{PhysRevB.65.165113,PhysRevB.74.174423,PhysRevB.87.125127} Note that all spin liquid states constructed here are so-called $Z_2$ spin liquids due to the condensation of boson bilinears $\hat{A}_{ij}$ and $\hat{t}^\gamma_{ij}$, which reduces the gauge symmetry from $U(1)$ to $Z_2$. 

Following Ref.~[\onlinecite{PhysRevB.87.125127}] we have preformed a PSG classification of all weakly symmetric ans\"atze (see Appendix \ref{appendix:WSansatz} for details). In the following we will only consider strictly symmetric, time-reversal invariant mean-field ans\"atze, however. Incidentally, there is just one ansatz in this class on the triangular lattice. It directly (i.e.~non-projectively) incorporates translation, point group and time reversal symmetries. This ansatz corresponds to the well-known zero-flux state in the Heisenberg limit and has the following properties.\cite{PhysRevB.45.12377,PhysRevB.74.174423,PhysRevB.87.125127} 
First of all, all bond parameters are real due to time reversal symmetry. Second, translation- 
and rotation symmetries ensure that the expectation values $A_{ij}=A$ and $B_{ij}=B$ are equal on all bonds. Note, however, that the $A_{ij}=-A_{ji}$ have a direction, which we choose such that $A$ is positive if it points in one of the three directions given by $\vec{a}_\gamma$.
Lastly, let's examine the action of three-fold rotations $C_3$ around the $\vec{z'}$ axis on the triplet bond operators $\hat{t}^\gamma$ (when using only two coordinates, they are expressed in the lattice coordinate system $(r_1,r_2)=r_1 \vec{a_x}+r_2 \vec{a_y}$). Under $C_3$ rotations the Schwinger boson operators transform as
\begin{align}
 C_3 \circ \begin{bmatrix}
b_{(r_1,r_2) \uparrow}^\dagger\\ b_{(r_1,r_2) \downarrow}^\dagger
\end{bmatrix} = 
e^{-\frac{i }{2 \sqrt{3}} \frac{2\pi}{3}(\sigma^x+\sigma^y+\sigma^z)}
\begin{bmatrix}
b_{(-r_2,r_1-r_2)\uparrow}^\dagger\\ b_{(-r_2,r_1-r_2)\downarrow}^\dagger
\end{bmatrix}. 
\end{align}
Accordingly, the triplet operators transform as
\begin{align}
C_3 \circ C_3(\hat{t}^x_{(0,0)(1,0)})=C_3(\hat{t}^y_{(0,0)(0,1)})=\hat{t}^z_{(0,0)(-1,-1)}.
\end{align}
Combined with the fact that $\hat{t}^\gamma_{ij}=\hat{t}^\gamma_{ji}$ it follows that the triplet bond parameters are all equal $\langle \hat{t}^\gamma_{ij}\rangle =t$ in a rotationally invariant state, as expected. 

To diagonalize the quadratic Hamiltonian, we first preform the Fourier transformation
\begin{align}
b_{\vec{r_i}\alpha}= \sum_\vec{k} e^{- i \vec{k} \vec{r_i}} b_{\vec{k} \alpha},
\end{align}
where $\vec{k}$ is summed over the first Brillouin zone. 
Using the pseudo-spinor notation 
$\Psi_\vec{k}=(b_{\vec{k}\uparrow },b_{\vec{k}\downarrow },b_{\vec{-k}\uparrow }^\dagger b_{\vec{-k}\downarrow }^\dagger)^T$, $H_\text{MF}$ reads
\begin{align}
\frac{1}{N} H_\text{MF}&=\frac{1}{N}\sum_\vec{k} \psi_\vec{k}^\dagger H_\vec{k} \psi_\vec{k} + 3 J_H |A|^2+ 3 J_K |t|^2 \nonumber \\
&- 3 (J_H+J_K) |B|^2 - \lambda (1+2S),
\label{eq:hmf0}
\end{align}
with the detailed form of the $4\times4$ matrix $H_\vec{k}$ given in Appendix \ref{appendix:HMF}. $H_\vec{k}$ can be diagonalized by a Bogoliubov transformation using a $SU(2,2)$ matrix $P_\vec{k}$ defined via
\begin{subequations}
\begin{align}
\Psi_\vec{k}&= P_\vec{k} \Gamma_\vec{k},\\
\Gamma_\vec{k}&=(\gamma_{\vec{k}1 },\gamma_{\vec{k}2 },\gamma_{\vec{-k}1 }^\dagger \gamma_{\vec{-k}2 }^\dagger)^T,\\
P_\vec{k}^\dagger H_\vec{k} P_\vec{k}&= \Omega_\vec{k} \ ,
\end{align}
\end{subequations}
where the $\gamma_{\vec{k}i}$ operators describe bosonic Bogoliubov quasiparticles, i.e.~bosonic spinons carrying spin 1/2.  $\Omega_\vec{k}$ is a diagonal matrix with two doubly degenerate eigenvalues, which take the form 
\begin{widetext}
\begin{subequations}
\begin{align}
\omega_{\pm}(\vec{k})&=\frac{1}{2} \sqrt{
\begin{aligned}
& \big(\lambda+(J_H+J_K)\tilde{B} \big)^2-J_H^2 \tilde{A}^2 -J_K^2 t^2 (\cos^2 k_1+\cos^2 k_2+\cos^2 k_3)\\
& \pm 2|J_H J_K \tilde{A}| \sqrt{t^2 (\cos^2 k_1+\cos^2 k_2+\cos^2 k_3)}
\end{aligned}}
\label{eq:dispersion},\\
\tilde{A}&=A(\sin(k_1)+\sin(k_2)+\sin(k_3)), \qquad
\tilde{B}=B (\cos(k_1)+\cos(k_2)+\cos(k_3)),
\end{align}
\label{eq:disp}
\end{subequations}
corresponding to half of the spinon excitation energy.
Above we defined $k_1=\vec{k} \cdot \vec{a_x}$, $k_2=\vec{k} \cdot \vec{a_y}$ and $k_3=-k_1-k_2$.
To determine the self consistent bond parameters, we compute the saddle points of ground state energy in the thermodynamic limit
\begin{align}
\frac{E_{GS}}{N}&= \int_{BZ} \frac{\text{d}^2k}{\text{Vol}_{BZ}} \big( \omega_+(\vec{k})+\omega_-(\vec{k}) \big)+ 3 \big(J_H |A|^2+ J_K |t|^2 -  (J_H+J_K) |B|^2 \big) - \lambda (1+2S) \ .
\end{align}
\end{widetext}
We used the Cuhre numerical integration routine from the Cuba package\cite{Hahn200578} to evaluate the above integral numerically. The relevant saddle point was found by maximizing $E_{MF}/N$ with respect to $B$ and $\lambda$ and minimizing with respect to $A$ and $t$.

\label{SSS Factor}
In order to characterize the SBMFT solutions for different spin liquid states, we compute the static spin structure factor, which can be measured directly in neutron scattering experiments. It is given by the equal-time spin-spin correlation function
\begin{align}
S(\vec{q})&=\frac{1}{N}\sum_{i,j} \langle \vec{S_i} \cdot \vec{S_j}  \rangle  e^{i \vec{q} (\vec{r_i}-\vec{r_j})} \ .
\end{align}
In addition we compute off-diagonal elements of the spin correlation tensor
\begin{align}
S^{c d}(\vec{q})&=\frac{1}{N}\sum_{i,j}  \langle S_i^c S_j^d \rangle  e^{i \vec{q} (\vec{r_i}-\vec{r_j})} \ .
\label{eq:corrtensor}
\end{align}
In order to calculate $S^{c d}(\vec{q})$ we express the spin operators in terms of Schwinger bosons, 
\begin{align}
S^{c d}(\vec{q})
&=\frac{1}{4 N} \sum_{\vec{k},\vec{k''}} \langle b_{\vec{k} \alpha}^\dagger \sigma_{\alpha \beta}^c b_{(\vec{k+q}) \beta}  b_{\vec{k''} \gamma}^\dagger \sigma_{\gamma \delta}^d b_{(\vec{k''-q}) \delta} \rangle ,
\label{eq:ssf2}
\end{align}
and use the Bogoliubov transformation matrix $P_\vec{k}$ to evaluate the expectation value in terms of the Bogoliubov quasiparticle operators using Wick's theorem. We get contributions $\sim\delta_{\vec{-k},\vec{k''}}$, $\delta_{\vec{k''},\vec{k+q}}$ as well as $\sim \delta_{\vec{q},0}$. The latter is proportional to the expectation value of the spin operator and thus vanishes in the spin liquid state. 

\begin{figure*}
\centering
    \includegraphics[width= 0.6\textwidth]{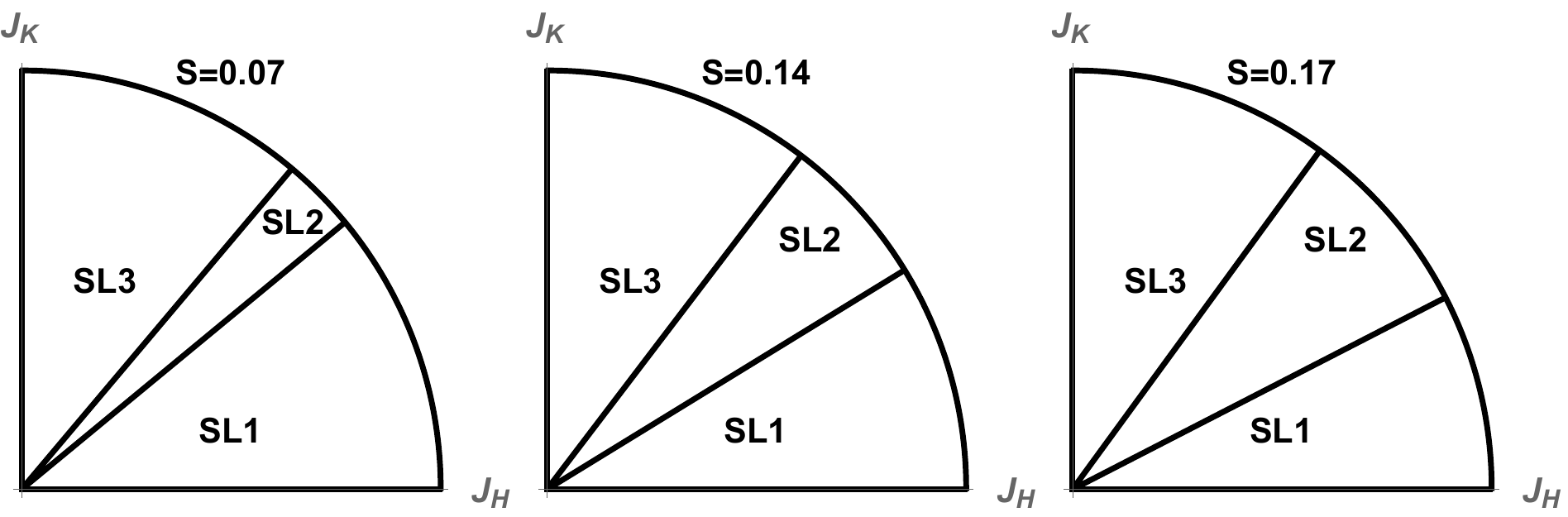}
\caption{Phase diagrams as function of the Heisenberg- and Kitaev couplings $J_H=\cos \psi$ and $J_K=\sin \psi$ for three different values of spin $S$. SL1, SL2 and SL3 denote three distinct spin liquid phases, separated by continuous phase transitions.}
\label{fig:PhaseD}
\end{figure*}

Since there are many Wick contractions, we used the Mathematica package SNEG for symbolic calculations with second-quantization-operator expressions.\cite{Zitko20112259}
The final expressions have the form:
\begin{align}
S^{c d}(\vec{q})&=
\frac{1}{\text{Vol}_k} \int \text{d}^2 k f^{c d}(P_\vec{k},P_\vec{k+q}),
\end{align}
where $f^{c d}(P_\vec{k},P_\vec{k+q})$ is a complicated function of around 50 terms consisting of the elements of the transformation matrices $P_\vec{k}$ and $P_\vec{k+q}$ which we do not state here explicitly.

\section{RESULTS}
\label{sec3}
\subsection{Self-consistent Mean-field Parameters}

\begin{figure*}
\centering
 \includegraphics[width= \textwidth]{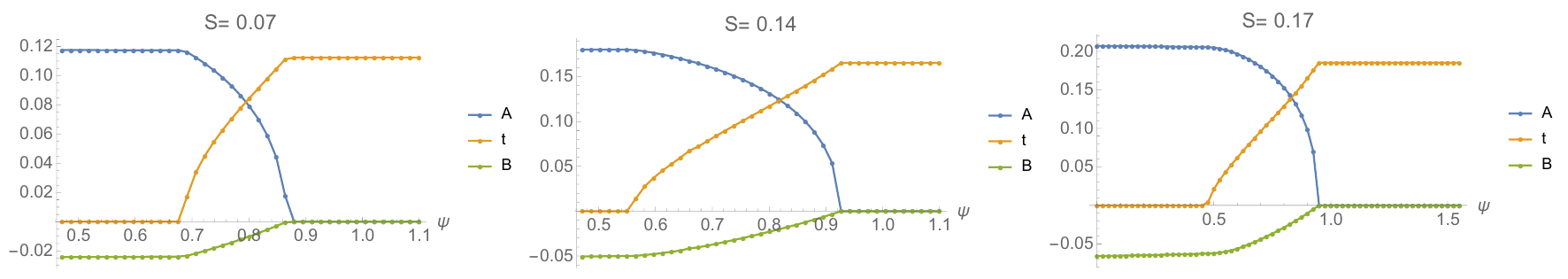}
\caption{(Color online) Saddle point values of the self-consistent mean-field parameters $A$,$t$ and $B$ as function of $\psi=\arctan (J_K/J_H)$, shown for three values of spin $S$. In all three cases the gap for spinon excitations is finite and the ground-state is thus a $Z_2$ spin liquid. 
}
\label{fig:SaddlePT}
\end{figure*}

\begin{figure}
\centering
    \includegraphics[width=\columnwidth]{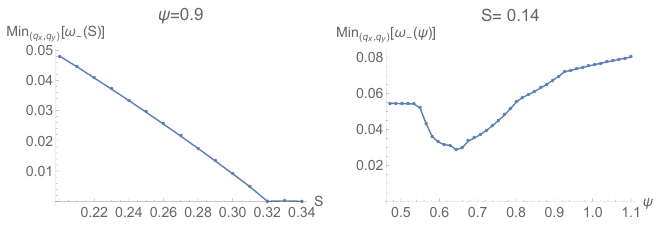}
\caption{
(Color online) Left: spinon gap as function of spin size $S$ at $\psi=0.9$. The gap closes at $S \approx 0.32$, leading to a magnetically ordered state for $S>0.32$.
Right: spinon gap as function of $\psi$ for $S=0.14$. Note that the gap remains finite throughout the continuous transitions between the three different $Z_2$ spin liquid phases.
}
\label{fig:gapC}
\end{figure}

In the following we restrict our discussion to the interesting regime where both, the Heisenberg- and the Kitaev coupling are antiferromagnetic $J_H>0$, $J_K>0$. We identify three distinct $Z_2$ spin liquid phases, denoted by SL1, SL2 and SL3, at sufficiently small spin $S$ and the corresponding phase diagrams for three different values of $S$ are shown in Figure \ref{fig:PhaseD}.
The saddle-point values of the self-consistent mean-field parameters $A$, $t$ and $B$ are shown as function of $\psi = \arctan J_K/J_H$ in Fig.~\ref{fig:SaddlePT} for the same three values of $S$. In SL1 the mean-field parameters $A$ and $B$ are nonzero, in SL2 all three parameters are nonzero, whereas only the parameter $t$ is different from zero in SL3. Note that the parameters vanish continuously at the transition between the three different $Z_2$ spin liquid phases, indicating the presence of two second-order quantum phase transitions as function of $J_K/J_H$.

Interestingly, the spinon gap remains finite at the two continuous phase transitions, as shown in Fig.~\ref{fig:gapC}. This points to an unconventional type of quantum phase transition between different symmetry enriched topological (SET) phases, which to our knowledge has not been discussed in the literature so far. Note that these continuous phase transitions connect states which do not differ in their PSG. Indeed, transitions between SET phases corresponding to different PSG's are expected to be discontinuous, i.e.~first order transitions (an example is discussed in Appendix~\ref{appendix:WSansatz}). 

Note that within the SBMFT description the SL1 as well as the SL3 phase have an emergent $SU(2)$ symmetry with a doubly degenerate spinon band. By contrast, in the SL2 phase this $SU(2)$ symmetry is broken and the degeneracy of the two spinon bands is lifted since both $A$ and $t$ are non-zero, as can be seen from Eq.~\eqref{eq:disp}. 

Finally we mention that the extension of the SL2 phase in parameter space is enlarged by increasing the spin size $S$. However, if the spin $S$ is increased beyond $S\gtrsim0.2$, the spinon gap closes and we get a Bose-Einstein condensate for some values of $\psi$, corresponding to a magnetically ordered state. In Fig.~\ref{fig:gapC} we plot the spinon gap as function of $S$ for $\psi=0.9$, where one can see that the gap closes continuously and magnetic order sets in beyond $S=0.32$. 

For an in-depth characterization of the three different spin liquid phases, we will focus our following discussion on six saddle points at $S=0.14$ with parameters shown in Table \ref{tab:points}. 
\vspace{.5cm}

\begin{table}
    \centering
   \begin{tabular}{|c|c|c|c|c|c|c|}
\hline
 Phase&$\psi$  & A & t & B & $\lambda$  & $E_{\text{MF}}$ \\
 \hline
SL1& 0 & 0.181 & 0 & -0.0528 & 0.4025 & -0.0901 \\
SL2& 0.6 & 0.176 & 0.0389 & -0.0476 & 0.303 & -0.0701 \\
SL2& 0.8 & 0.132 & 0.117 & -0.023 & 0.281 & -0.0635 \\
SL2& 0.85 & 0.109 & 0.135 & -0.0147 & 0.276 & -0.0634 \\
SL2& 0.9 & 0.0687 & 0.154 & -0.0055 & 0.272 & -0.0645 \\
 SL3&$\frac{\pi }{2}$ & 0 & 0.165 & 0 & 0.338 & -0.082 \\
 \hline
 \end{tabular}
    \caption{Six saddle points for different $\psi=\arctan J_K/J_H$ at $S=0.14$, used for further characterization of the three different spin liquid phases. $A$, $t$, and $B$ denote the expectation values of the corresponding bond operators, $\lambda$ is the value of the Lagrange multiplier and $E_\text{MF}$ denotes the mean-field ground state energy per spin.}
    \label{tab:points}
\end{table}
\subsection{Spinon Dispersion}

Even though the spinon dispersion is not gauge invariant and thus not a directly observable quantity, it is nevertheless an interesting object. In particular, the minima of the spinon dispersion determine the structure of the spinon condensate in the magnetically ordered phase, which allows us to determine the position of Bragg peaks in the spin structure factor. 

\begin{figure*}
\centering
\includegraphics[width= 0.95\textwidth]{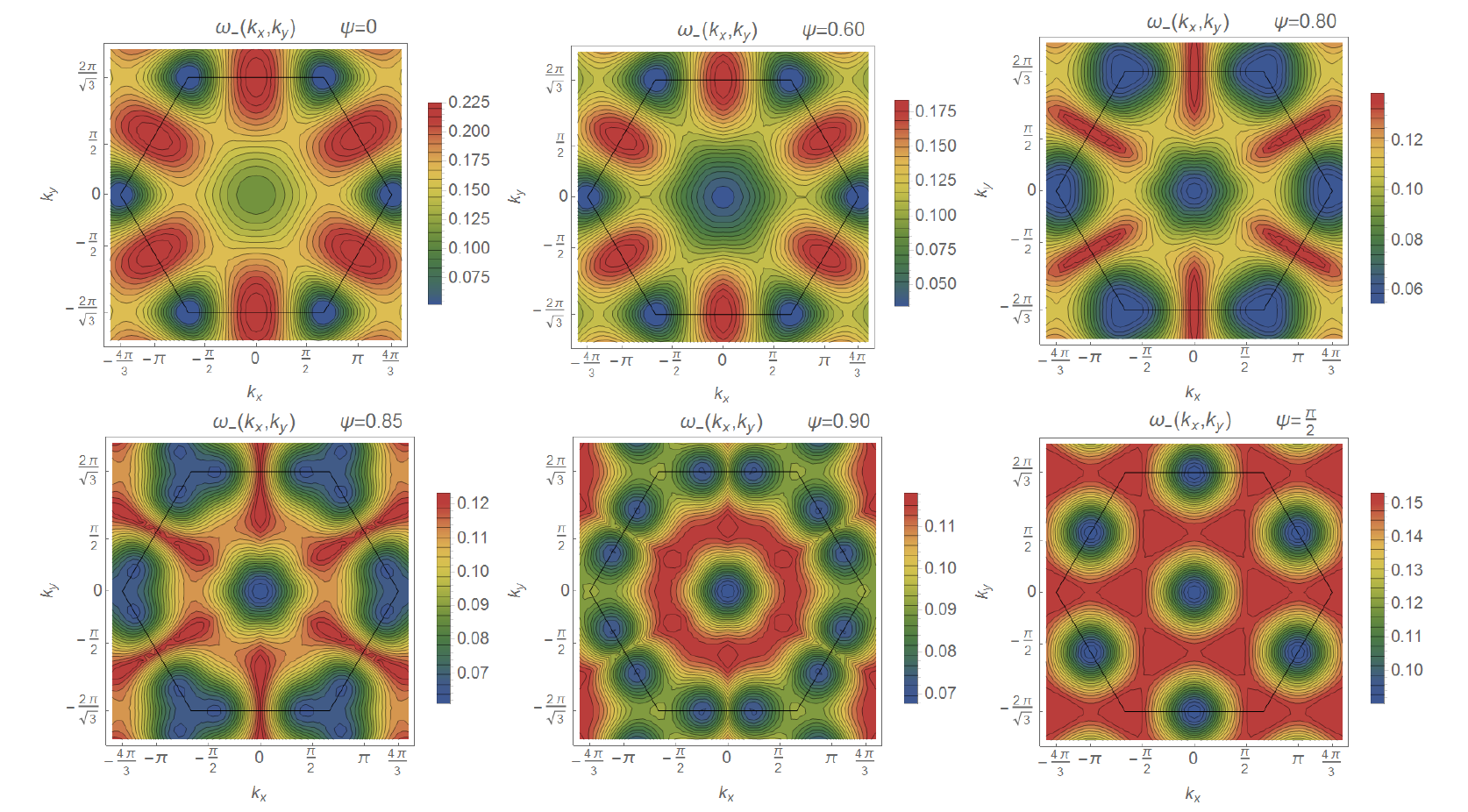}
\caption{(Color online) Lower branch of the spinon dispersion $\omega_-(\vec{k})$ from Eq.~\eqref{eq:dispersion} for the six different values of $\psi$ shown in Table \ref{tab:points}. The black hexagon marks the boundary of the first Brillouin zone. Note that by increasing $\psi$ the minima of the dispersion move from the $K$ points at the corners of the first Brillouin zone in SL1, through incommensurate momenta in SL2 to the $M$ points at the middle of the Brillouin zone edges as well as to the $\Gamma$ point ($\vec{q}=0$) in SL3 (see main text for a discussion).}
\label{fig:dispersions}
\end{figure*}

The lower energy branch $\omega_-$ of the spinon dispersion in Eq.~\eqref{eq:disp} is shown in Fig.~\ref{fig:dispersions} for the six saddle points listed in Table \ref{tab:points}. The minima of the dispersion in the SL1 phase are at the $K$ points (i.e.~the corners) of the first Brillouin zone, which is consistent with previous results.\cite{PhysRevB.45.12377} Condensing spinons at the $K$ points for larger values of $S$ indeed leads to the well known 120-degree order, as will be shown in Sec.~\ref{sec:classSL1}. By contrast, the degenerate dispersion minima in the SL3 phase appear at the $M$ points in the middle of the edges of the first Brillouin zone, as well as at the $\Gamma$ point at $\vec{q}=0$. 

In the SL2 phase close to the transition to SL1, the dispersion minima remain at the $K$ points of the first Brillouin zone. Upon increasing $\psi$ the minima start to move to incommensurate momenta along the edge of the first Brillouin zone. For some values of $\psi$ and $S$ the global minimum jumps to zero momentum, however. We believe that this is an artifact of the mean-field approximation, where the $B$ fields are likely overestimated. Indeed, changing the value of $B$ by a few percent already shifts the absolute minimum to the previously mentioned incommensurate momenta at the Brillouin zone edges. This is important, because a spinon condensate at $\vec{q}=0$ alone would lead to a ferromagnetically ordered state, which is not expected for antiferromagnetic couplings $J_H$ and $J_K$.

Our mean-field analysis thus seems to be not always reliable to accurately determine the position of the dispersion minima in the SL2 phase. Indeed, we will argue later on that dispersion minima at incommensurate wave vectors along the Brillouin zone edges are in accordance with the expected magnetic order in the classical limit.

\subsection{Static Spin Structure Factor}

\begin{figure*}
\centering
\includegraphics[width= 0.95\textwidth]{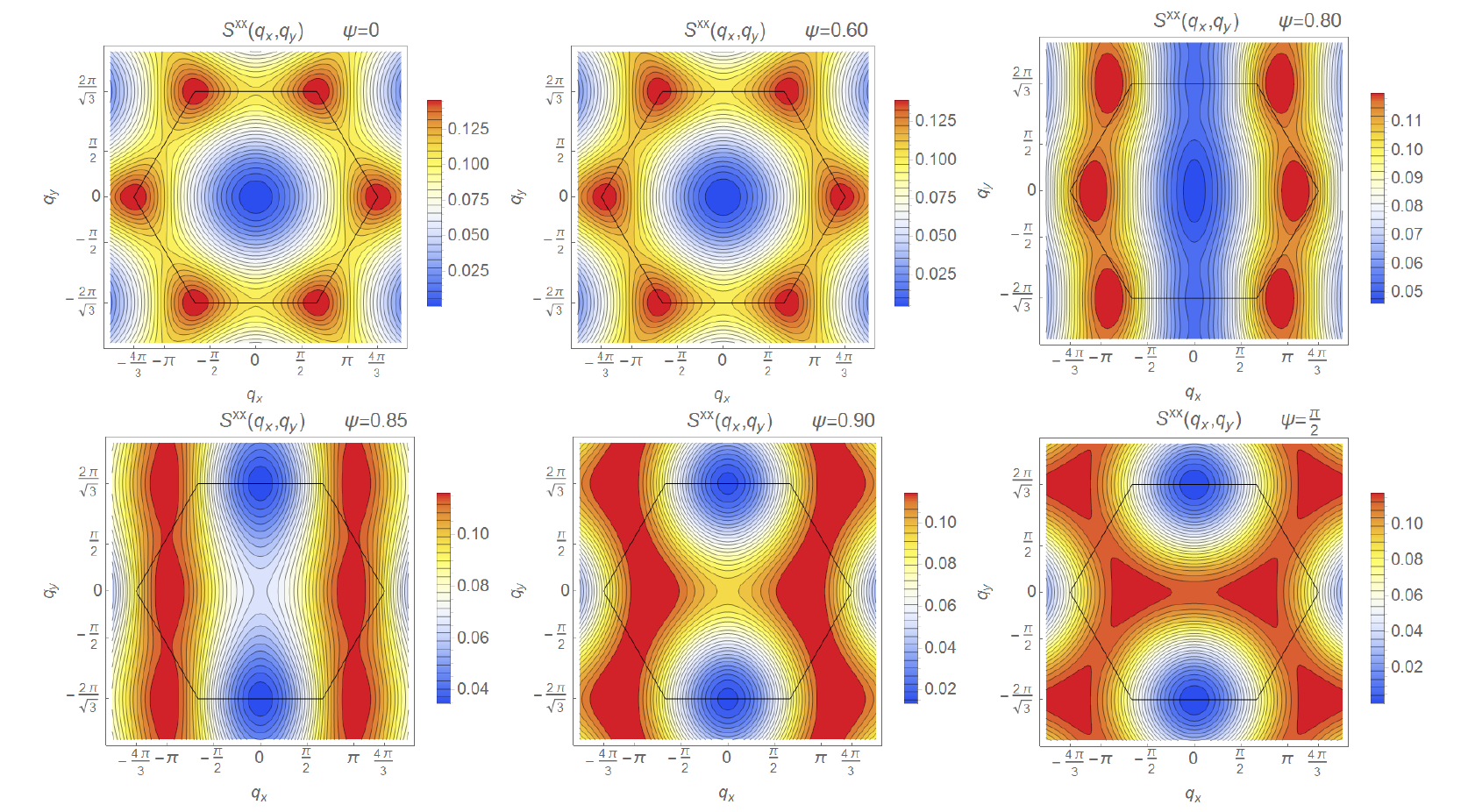}
\caption{(Color online) Diagonal component $S^{xx}(\vec{q})$ of the spin correlation tensor Eq.~\eqref{eq:corrtensor} for the six points shown in Table \ref{tab:points}. Note that the other diagonal components $S^{yy}(\vec{q})$ and $S^{zz}(\vec{q})$ can be obtained simply from $S^{xx}(\vec{q})$ by $\pm\pi/3$ rotations around the $\Gamma$ point. Note that that the maxima move to incommensurate momenta in the SL2 phase.}
\label{fig:Sxx}
\end{figure*}

The static spin structure factor can be measured directly in neutron scattering experiments. Note that due to the presence of fractionalized spinon excitations in a spin liquid, the structure factor exhibits a broad two-spinon continuum, rather than sharp Bragg peaks as in a magnetically ordered phase.

In Fig.~\ref{fig:Sxx} we show the results for the $xx$-component $S^{xx}(\vec{q})$ of the structure factor for the six different saddle points listed in Tab.~\ref{tab:points}. While $S^{xx}(\vec{q})$ is symmetric under six-fold rotations around the $\Gamma$ point and is peaked at the $K$ points in the SL1 phase, the maxima change their position and move to incommensurate momenta upon increasing $\psi$ in the SL2 phase, where the six-fold rotation symmetry is lost as well. Finally, in the SL3 phase $S^{xx}(\vec{q})$ is peaked at $\vec{q}=\pm(2 \pi/3,0)$. We note that equivalent peaks have been observed in related Kitaev-type models in Ref.~[\onlinecite{PhysRevB.86.155127}].

The other two diagonal elements of the spin correlation tensor $S^{yy}(\vec{q})$ and $S^{zz}(\vec{q})$ can be obtained from $S^{xx}(\vec{q})$ simply through a rotation by $\pm \frac{\pi}{3}$ around the $\Gamma$ point in all phases, and are thus not shown explicitly.

\begin{figure*}
\centering
\includegraphics[width= 0.95\textwidth]{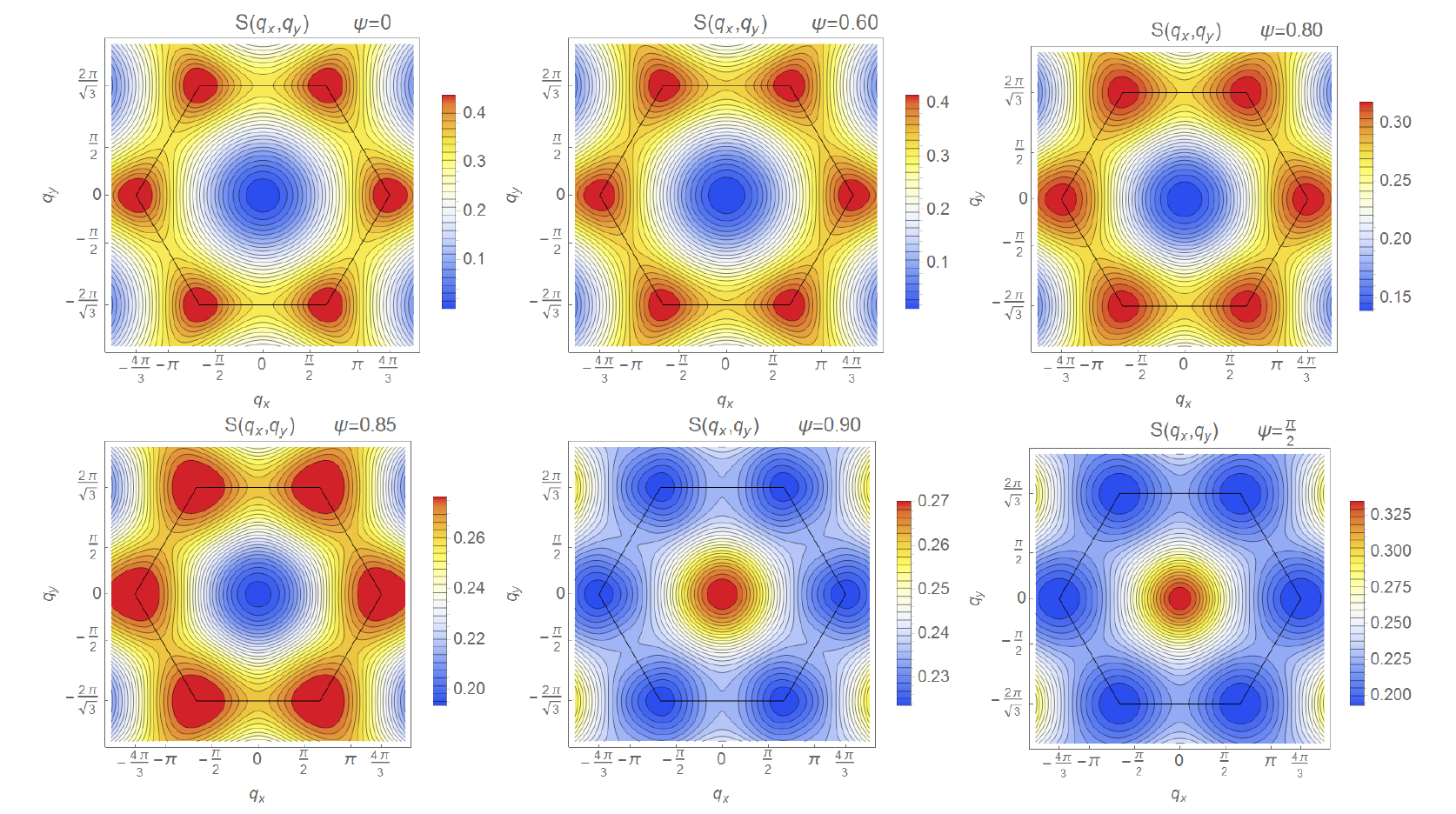}
\caption{(Color online) Total static spin structure factor $S(\vec{q})$ for the six saddle points shown in Table \ref{tab:points}. The peaks of $S(\vec{q})$ stay at commensurate momenta, in contrast to the maxima of the diagonal components $S^{aa}(\vec{q})$.}
\label{fig:S}
\end{figure*}

The total static spin structure factor is just the sum of the diagonal elements
\begin{align}
    S(\vec{q})=S^{xx}(\vec{q})+S^{yy}(\vec{q})+S^{zz}(\vec{q}),
\end{align}
and is shown in Fig.~\ref{fig:S}. Here, the peaks stay at commensurate momenta in all phases, in contrast to the peaks of the individual diagonal components $S^{aa}(\vec{q})$.

We also note that even though the total magnetization $\langle \sum_i \vec{S}_i \rangle=0$ vanishes in all spin liquid ground states per construction, the variance of the total spin $\langle (\sum_i \vec{S}_i )^2 \rangle$ is non-zero in the SL2 and SL3 phases, as evidenced by the fact that the structure factor doesn't vanish at the $\Gamma$ point (see Fig.~\ref{fig:S}). This is due to the fact that the ground-state is not a total spin singlet in SL2 and SL3, because the triplet fields $\hat{t}_\gamma$ acquire a non-zero expectation value.

\begin{figure*}
\centering
\includegraphics[width= 0.95\textwidth]{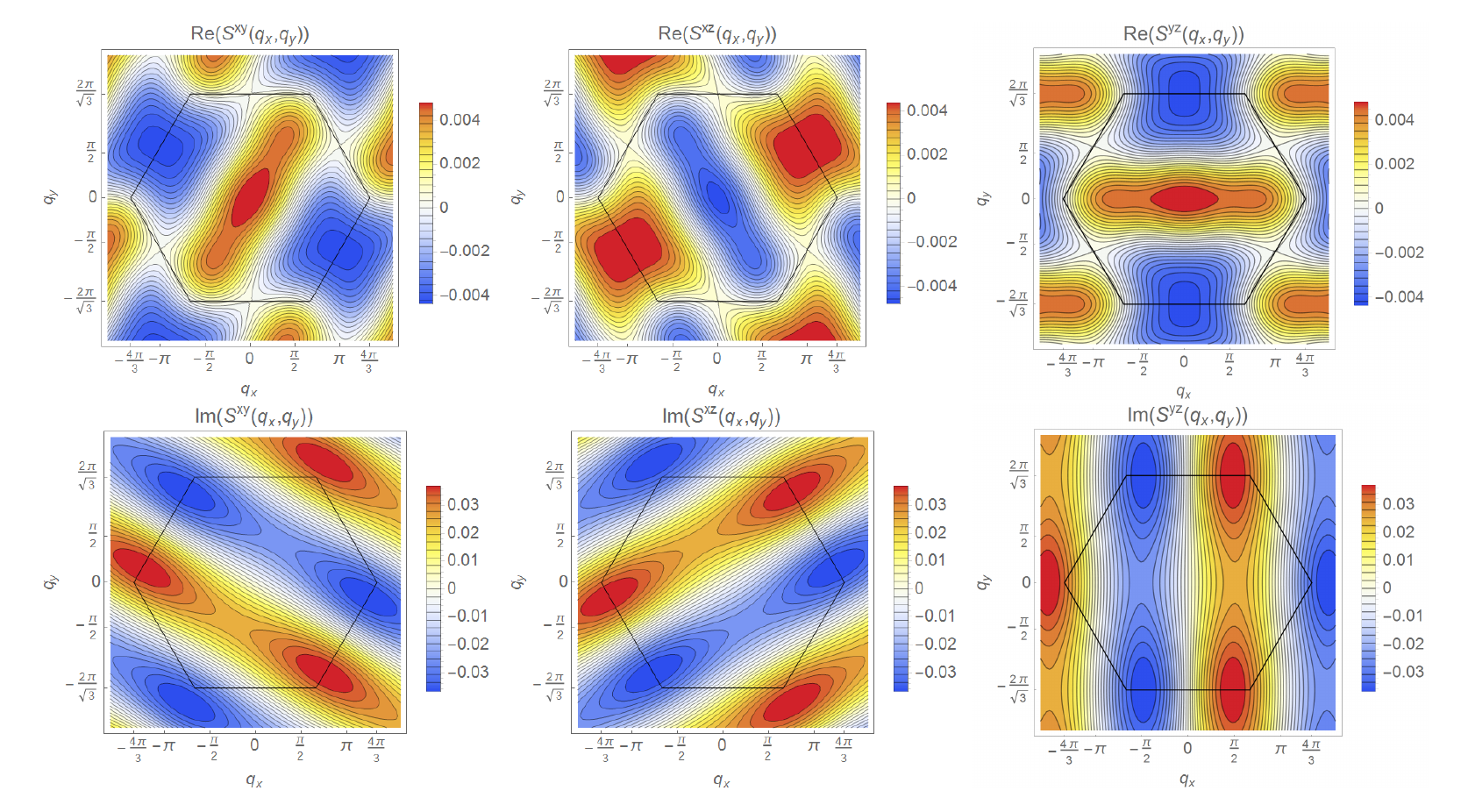}
\caption{(Color online) Real and imaginary part of the off-diagonal elements $S^{ab}(\vec{q})$ computed for one point in the SL2 phase ($\psi=0.85$).  By contrast, these off-diagonal elements vanish identically in the SL1 and SL3 phases. Note that $S^{ab}(\vec{q})$ cannot be diagonalized simultaneously for all momenta $\vec{q}$.}
\label{fig:Sab}
\end{figure*}

The defining observable difference between the SL2 phase and the other two phases is that the off-diagonal elements of the spin correlation tensor $S^{ab}(\vec{q})$ defined in Eq.~\eqref{eq:corrtensor} \emph{vanish} in SL1 and SL3, but are \emph{non-zero} in SL2, indicating correlations between different spin components.  In Fig.~\ref{fig:Sab} we show the real- and imaginary parts for the off-diagonal components of $S^{ab}(\vec{q})$ for one particular point in the SL2 phase ($\psi=0.85$).
Even though $S^{ab}(\vec{q})$ can be diagonalized for any given momentum $\vec{q}$, i.e.~one can find a frame where any two spins are collinear, it is important to note that $S^{ab}(\vec{q})$ is \emph{not} diagonal in the same frame for \emph{all} momenta $\vec{q}$. This is in marked contrast to SL1 and SL3.

\subsection{Spinon Condensation and Classical Limit}

One advantage of the SBMFT approach over fermonic slave-particle approaches is that SBMFT simply allows for the description of magnetically ordered states through spinon condensation.
As mentioned earlier, the spinon gap decreases upon increasing the spin length $S$ and closes at a discrete set of momenta in the Brillouin zone at some critical value $S_c$, as shown in Figure \ref{fig:gapC}. 
Further increasing $S$ adds bosons in the zero modes and a Bose-Einstein condensate forms, resulting in a non-zero expectation value of the spin operator $\langle \vec{S}_i \rangle$ and we get a magnetically ordered state.
The size of the ordered magnetic moment (i.e.~the density of Bose-Einstein condensate) is determined by the saddle point conditions.

\subsubsection{Classical Limit in SL1 phase}
\label{sec:classSL1}
The magnetic order parameter obtained by condensing spinons in the SL1 phase has been discussed already in the context of the Heisenberg model on the triangular lattice by Sachdev [\onlinecite{PhysRevB.45.12377}] as well as by Wang and Vishwanath [\onlinecite{PhysRevB.74.174423}]. We briefly review these results here.

In the SL1 phase the zero modes appear at the corners ($K$ points) of the first Brillouin zone. The corresponding two inequivalent momenta are $\vec{k_c}=\pm(4 \pi/3,0)$. The structure of the magnetic order parameter is determined by the eigenvectors of the matrix $\tau^z H_\vec{k_c}$, i.e.~the columns of the Bogoliubov rotation matrix $P_\vec{k_c}$. At $\vec{k_c}$ the eigenvalues of $\tau^z H_\vec{k_c}$ are doubly degenerate, so we get two orthogonal eigenvectors
\begin{align}
    \psi_1(\vec{k_c})&=(i,0,0,1)^T, \nonumber\\
    \psi_2(\vec{k_c})&=(0,-i,1,0)^T.
\end{align}
The spinon condensate thus has the form
\begin{align}
   \langle \begin{pmatrix}
b_{\vec{k_c}\uparrow} \\ b_{\vec{k_c} \downarrow}\\
b_{\vec{-k_c}\uparrow}^\dagger \\ b_{\vec{-k_c} \downarrow}^\dagger
\end{pmatrix} \rangle=
s_1 \psi_1(\vec{k_c})+s_2 \psi_2(\vec{k_c})=
\begin{pmatrix}i s_1\\-i s_2 \\s_2\\s_1
\end{pmatrix}
\end{align}
where $s_1$ and $s_2$ are complex constants and only $|s_1|^2+|s_2|^2$ is fixed by the size of spin $S$. In real space the spinon condensate is given by
\begin{align}
  x \equiv \langle \begin{pmatrix}
b_{\vec{r}\uparrow} \\ b_{\vec{r} \downarrow}
\end{pmatrix} \rangle
=
\begin{pmatrix}
i c_1 e^{i \vec{k_c}\vec{r}}-i c_2 e^{-i \vec{k_c}\vec{r}}\\
 c_2^* e^{i \vec{k_c}\vec{r}}+ c_1^* e^{-i \vec{k_c}\vec{r}}
\end{pmatrix}
\end{align}
where we redefined the constants as $c_1= s_1$ and $c_2=i s_2^*$ to match the notation in Ref.~[\onlinecite{PhysRevB.74.174423}].

The ordered magnetic moment can easily be calculated from $x$ as $\vec{S(\vec{r})}=\frac{1}{2}x^\dagger \boldsymbol{\sigma} x $ and corresponds to the well known co-planar $120^\circ$ order, as expected in the Heisenberg limit.\cite{PhysRevLett.82.3899} The freedom to choose $c_1$ and $c_2$ is just a consequence of the global $SU(2)$ symmetry, which allows to arbitrarily rotate the plane in which the magnetic moments lie.
\subsubsection{Classical Limit in SL3 phase}

\begin{figure}
\centering
\includegraphics[width= \columnwidth]{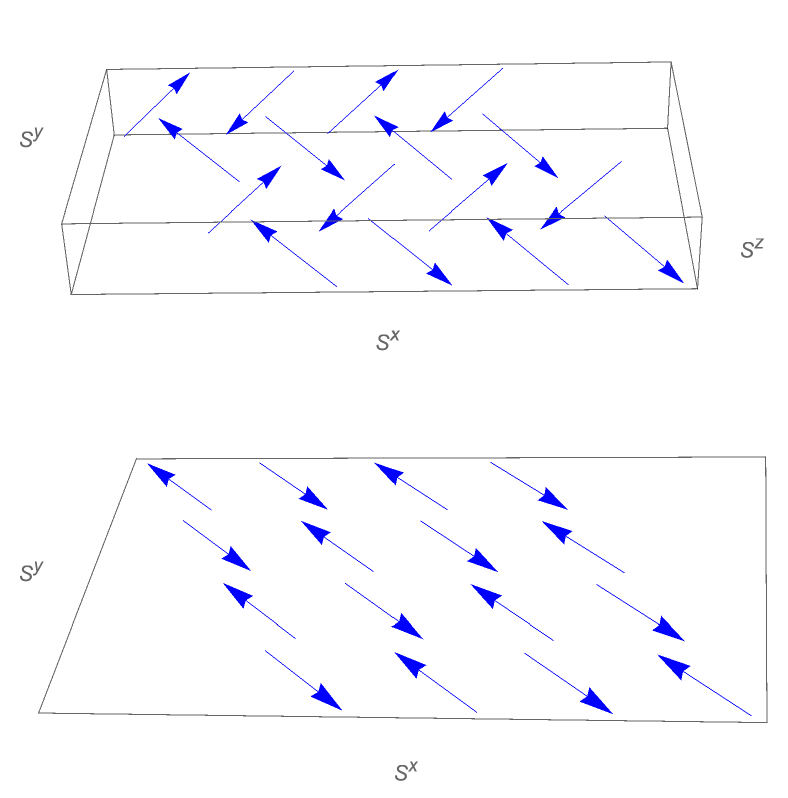}
\caption{(Color online) Sketches of the magnetic order obtained by condensing spinons in the SL3 phase. Top: same condensate density at all four zero modes. Bottom: condensate density vanishes at two of the four zero modes. For illustrative purposes we rotated the spin components $S^\gamma$ to the primed lattice coordinate system shown in Fig.~\ref{fig:triangular}. Note the antiferromagnetically ordered chains along one of the three directions of the triangular lattice, as expected from the classical limit at the Kitaev point.}
\label{fig:SL3spins}
\end{figure}

Determining the magnetic order parameter in the large $S$ limit of the SL3 phase is more cumbersome, because now we have four nonequivalent zero modes, corresponding to the three non-equivalent $M$ points, as well as the $\Gamma$ point in the Brillouin zone as shown in the last panel of Fig.~\ref{fig:dispersions}. Moreover, the Hamiltonian is no longer $SU(2)$ symmetric. The four inequivalent momenta of the zero modes are given by
\begin{align}
  \vec{k_{c0}}&=(0,0) && \vec{k_{c1}}=(0,\frac{2\pi}{\sqrt{3}})
  \nonumber \\
  \vec{k_{c2}}&=(\pi,\frac{\pi}{\sqrt{3}}) && \vec{k_{c3}}=(\pi,-\frac{\pi}{\sqrt{3}}).
\end{align}
Again each eigenvalue is doubly degenerate, so we get two eigenvectors for every zero mode
\begin{subequations}
\begin{align}
  \psi_{01}&=\left(
 \frac{1-i}{\sqrt{6}} ,\frac{i}{\sqrt{6}},
 \frac{1}{\sqrt{2}},0 \right)^T, 
 \\
 \psi_{02}&=\left(
 \frac{i}{\sqrt{6}} ,\frac{1+i}{\sqrt{6}},0,
 \frac{1}{\sqrt{2}} \right)^T, 
 \\
  \psi_{11}&=\left(
 \frac{-1-i}{\sqrt{6}} ,\frac{-i}{\sqrt{6}},
 \frac{1}{\sqrt{2}},0 \right)^T, 
 \\
 \psi_{12}&=\left(
 \frac{-i}{\sqrt{6}} ,\frac{-1+i}{\sqrt{6}},0,
 \frac{1}{\sqrt{2}} \right)^T, 
\end{align}
\begin{align}
   \psi_{21}&=\left(
 \frac{1+i}{\sqrt{6}} ,\frac{-i}{\sqrt{6}},
 \frac{1}{\sqrt{2}},0 \right)^T, 
 \\
 \psi_{22}&=\left(
 \frac{-i}{\sqrt{6}} ,\frac{1-i}{\sqrt{6}},0,
 \frac{1}{\sqrt{2}} \right)^T, 
 \\
   \psi_{31}&=\left(
 \frac{-1+i}{\sqrt{6}} ,\frac{i}{\sqrt{6}},
 \frac{1}{\sqrt{2}},0 \right)^T, 
 \\
 \psi_{32}&=\left(
 \frac{i}{\sqrt{6}} ,\frac{-1-i}{\sqrt{6}},0,
 \frac{1}{\sqrt{2}} \right)^T .
\end{align}
\end{subequations}
Accordingly, the condensate takes the form
\begin{align}
   \langle \begin{pmatrix}
b_{\vec{k}\uparrow} \\ b_{\vec{k} \downarrow}\\
b_{\vec{-k}\uparrow}^\dagger \\ b_{\vec{-k} \downarrow}^\dagger
\end{pmatrix} \rangle
=
\sum_{i=0}^3 \left( c_{i1} \psi_{i1}+c_{i2} \psi_{i2} \right) \delta_{\vec{k},\vec{k_{ci}}}
\end{align}
In contrast to the SL1 case, we now have 8 complex constants $c_{ij}$ to determine the order parameter.
In order to reduce the number of independent constants we use the fact that the points $\vec{k_{ci}}$ and $-\vec{k_{ci}}$ are equivalent as they are connected by reciprocal lattice vectors. Self consistency of our description thus requires that 
$\langle b_{\vec{k_{ci}}\alpha} \rangle = \langle b_{\vec{-k_{ci}}\alpha}\rangle$ must hold, 
which gives four equations
\begin{align}
   c_{i1}^* (\psi_{i1})_3^* + c_{i2}^* (\psi_{i2})_3^* = 
 c_{i1} (\psi_{i1})_1 + c_{i2} (\psi_{i2})_1,
\end{align}
where $i$ goes from 0 to 3.
Taking these equations into account, there are still four complex parameters to be fixed in order to uniquely determine the magnetic order parameter. To do this, it would be necessary to consider interactions between the various condensate modes, which would require us to go beyond mean-field theory. In the following we circumvent this problem by making a few reasonable assumptions about the structure of the condensate. First, we demand a constant density of condensed bosons at each lattice site, which translates to the fact that the size of the ordered moment is the same on each lattice site (note that this assumption definitely holds in the classical limit $S \to \infty$).  Since the order is commensurate, we have only four inequivalent sites which gives four additional constraints. These fix the phases of the condensates at the zero modes. 
Second, we demand that the total magnetization vanishes. These conditions still do not fully determine the structure of the condensate, which is due to the fact that we still have a $D_{2h}$ symmetry in spin space at the Kitaev point.
Incorporating all constraints, there remains some freedom in choosing the amplitudes of the condensates at the various zero modes. Demanding that either all four or two modes have the same occupancy, we numerically solved the constraint equations and the resulting magnetic order is shown in Figure \ref{fig:SL3spins}. 
The ordered magnetic moment for the two above mentioned amplitude choices can be written as
\begin{subequations}
\begin{align}
   \vec{S}(n,m)&=\frac{1}{\sqrt{2}} \left( (-1)^{n+m},0,(-1)^{n} \right)^T \\
    \vec{S}(n,m)&=\frac{1}{\sqrt{2}} \left( (-1)^{n+m},(-1)^{n+m} ,0 \right)^T,
\end{align}
\end{subequations}
where the first expression corresponds to the case where the condensate density is equal in all four modes, whereas in the second expression the condensate density is zero for two modes and equal in the other two. Here, we labelled sites on the triangular lattice by the integer indices $n$ and $m$ via $\vec{r}= n \vec{a_x}+m \vec{a_y}=(n,m)$. Note that both solutions indeed correspond to degenerate ground states of the classical Kitaev model on the triangular lattice with the energy $E_0=-N J_K$.\cite{PhysRevB.93.104417}  Even though the full set of degenerate classical ground states also contains non-coplanar spin configurations, we only recover states with coplanar order, because we start from a real, time-reversal symmetric ansatz. In Appendix~\ref{appendix:Anstzes 4}  we show that degenerate non-coplanar states can be obtained in the large $S$ limit starting from a weakly symmetric ansatz which breaks time reversal symmetry.

\subsubsection{Classical Limit in SL2 phase}

Finally, we comment on the classical limit of the SL2 phase. Assuming that the zero modes are located at incommensurate momenta along the Brillouin zone edges as discussed above, we get six inequivalent condensate modes. In this case interactions between condensate modes are clearly important and it is no longer possible to determine the structure of the condensate and the magnetic order parameter within mean-field theory, because there are too many free parameters and too few constraints.  

\begin{figure}
\centering
    \includegraphics[width=0.55\columnwidth]{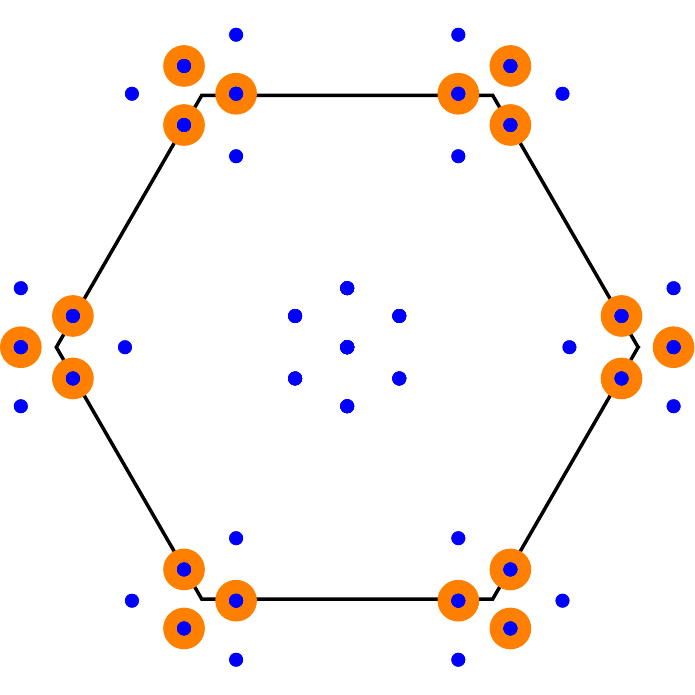}
\caption{(Color online) Schematic positions of magnetic Bragg peaks after spinon condensation in the SL2 phase. The discrete set of zero modes  at momenta $\vec{k_{ci}}$ (orange, large) where the spinon gap vanishes determine the possible positions of Bragg peaks in the static spin structure factor (blue, small), which are given by the differences $\vec{k_{ci}-k_{cj}}$ modulo reciprocal lattice vectors. The black hexagon marks the first Brillouin zone. }
\label{fig:shem}
\end{figure}

For this reason we restrict our discussion of the classical limit to the location of Bragg peaks in the static structure factor. 
As already mentioned, the spinon gap closes at a discrete set of momenta $\{\vec{k_{ci}}\}$, where i runs over all nonequivalent zero modes in the first Brillouin zone. As can be seen from Eq.~\eqref{eq:ssf2}, the possible positions of magnetic Bragg peaks in the spin structure factor are determined by the differences of the zero mode momenta $\vec{k_{ci}-k_{cj}}$ modulo reciprocal lattice vectors. The resulting Bragg peak positions in the SL2 phase are shown schematically in Figure \ref{fig:shem}. Note that we assumed here that the zero modes only appear at incommensurate momenta along the edges of the Brillouin zone (see discussion of the spinon dispersion above). 
Notably, the position of Bragg peaks on the Brillouin zone edges is in accordance with expectations for the Z2 vortex crystal phase, which is the ground state of the classical Heisenberg-Kitaev model on the triangular lattice for $J_H>0$ and $J_K>0$.\cite{PhysRevB.93.104417, PhysRevB.91.155135} 
Consequently, it appears as if the SL2 phase might represent a quantum disordered version of the classical Z2 vortex crystal.
 
However, we used a real ansatz and the corresponding spin liquid states are time reversal symmetric. In the classical limit (if it exists) we thus expect coplanar order,\cite{PhysRevB.87.125127} which is different from the non-coplanar order in the Z2 vortex phase.\cite{PhysRevB.93.104417} This is apparent by looking at the scalar spin chirality (defined for spins on an elementary triangle)
\begin{equation}
\chi_{ijk} = \vec{S}_i \cdot (\vec{S}_j \times \vec{S}_k) \sim \text{Im} (\hat{B}_{ij} \hat{B}_{jk} \hat{B}_{ki}) = 0 ,
\end{equation}
which vanishes in all time reversal invariant states considered here, but is non-zero in the Z2 vortex crystal.
This problem can be resolved by considering so-called weakly-symmetric ans\"atze, which obey lattice symmetries but are allowed to break time-reversal and parity symmetries.\cite{PhysRevB.87.125127} In this case some mean-fields obtain a complex phase and the scalar spin chirality is non-zero. We briefly discuss chiral bosonic states in Appendix~\ref{appendix:WSansatz}. 

Finally, we mention that the position of the other three Bragg peaks in the vicinity of the Brillouin zone corner coincides with the maxima of the three diagonal elements of the structure factor $S^{aa}(\vec{q})$ in the SL2 phase, which are thus expected to carry the largest weight.

\subsection{Phase Transitions}

The two continuous quantum phase transitions between the different $Z_2$ spin liquids have interesting properties. First of all, the excitation gaps of all elementary excitations remain finite throughout both transitions. Indeed, the spinon gap is always finite, as discussed in detail above. Besides spinon excitations a $Z_2$ spin liquid exhibits vortex excitations of an emergent $Z_2$ gauge field, which are gapped but carry no spin.\cite{PhysRevLett.66.1773,PhysRevB.44.2664,PhysRevB.62.7850,PhysRevB.84.094419} These so called visons are not accounted for in the SBMFT description. They are related to phase fluctuations of the mean fields, which are gapped due to the Higgs mechanism by condensing either the $\hat{A}_{ij}$ or the $\hat{t}_{ij}^\gamma$ bond operators. In our case the vison gap is finite throughout the transitions as well, because either the $A$ or the $t$ fields are condensed in any case. The question thus remains which excitation gap closes at the continuous transition between two different $Z_2$ spin liquid phases.

Considering the transition between SL1 and SL2, we see from Fig.~\ref{fig:SaddlePT} that the triplet amplitude $t$ vanishes at the critical point. Consequently collective fluctuations of the triplet fields $\hat{t}^\gamma_{ij}$ are gapless at the critical point. Since all other excitations are gapped, we can formally integrate out the spinons as well as the fields $A$ and $B$, which can be viewed as Hubbard-Stratonovich fields, and derive an effective action for the triplet fields $t$ alone. The form of the Lagrangian is severely constrained by gauge invariance and takes the form
\begin{equation}
L_\text{eff}=L_B+\sum_{i,j} \left( c_1 |t^\gamma_{ij}|^2 +c_2 |t^\gamma_{ij}|^4 \right) + \sum_\text{loops} t^{\gamma *}_{ij} t^\gamma_{jk}  t^{\beta *}_{k\ell} t^\beta_{\ell i} + \cdots
\end{equation} 
where $L_B$ is a Berry phase term involving temporal gradients of the fields $t^\gamma_{ij}(\tau)$, the dots denotes further gauge invariant terms on even loops and $c_{1,2}$ are real constants. Making the reasonable assumption that the Berry phase term only contains second order derivatives, there are only few potential universality classes for the critical point, depending on the detailed form of the fourth order term. Among the possible universality classes are the fixed point of the three dimensional $O(6)$ model, or a cubic fixed point.\cite{PhysRevB.61.15136} Similar arguments can be applied to study the transition between SL2 and SL3. We leave a detailed investigation of these phase transitions open for future study.

\section{Discussion and CONCLUSIONs}
\label{sec4}

We used Schwinger boson mean-field theory to construct $Z_2$ spin liquid ground states of the Heisenberg-Kitaev model on the triangular lattice in the regime where both, the Heisenberg as well as the Kitaev coupling are antiferromagnetic. Using a symmetric, time-reversal invariant mean-field ansatz we found three distinct spin liquid phases for sufficiently small spins $S$.
At the physical value $S=1/2$ the spinons are condensed in all phases and the ground-state has magnetic order.

The $SU(2)$ symmetric SL1 spin liquid phase shows up in the vicinity of the Heisenberg point and corresponds to the well known zero-flux state.\cite{PhysRevB.45.12377,PhysRevB.74.174423} It develops $120^\circ$ order in the classical limit, which agrees with previous results of the Heisenberg model on the triangular lattice. Note that SL1 is stable with respect to small Kitaev couplings $J_K$ in our approach. This is in contrast to numerical studies as well as to the behavior in the classical limit, where the $120^\circ$ order is unstable towards a $Z_2$ vortex crystal for infinitesimally small $J_K$. We emphasize, however,  that the region occupied by SL1 in the phase diagram shrinks as the spin $S$ is increased, which follows the correct trend that SL1 is unstable to small Kitaev couplings in the limit of large $S$.

Increasing the antiferromagnetic Kitaev coupling $J_K$ beyond a critical value we observe a second order transition from SL1 to a different spin liquid phase, dubbed  SL2, which has interesting properties. Here, the minima of the spinon dispersion appear at incommensurate momenta. 
An interesting feature of the SL2 phase is that the maxima of the diagonal elements of the spin correlation tensor $S^{ab}(\vec{q})$ are shifted away from the Brillouin zone corners. Moreover, the SL2 phase exhibits non-zero off-diagonal elements of $S^{ab}(\vec{q})$, indicating unconventional spin correlations. 
This is in marked contrast to the SL1 and SL3 phases, where off-diagonal spin correlations are absent.

In the classical limit of the SL2 phase certain Bragg peak positions coincide with Bragg peaks in the exotic $Z_2$ vortex crystal phase, which has been discussed in detail in the literature.\cite{PhysRevB.93.104417, PhysRevB.91.155135}  
It remains to be seen what type of magnetic order develops in the classical limit of the SL2 phase, however.
Using a time reversal symmetric ansatz gives rise to coplanar order in the classical limit (if it exists). To obtain a non-coplanar state, like the Z2 vortex crystal, one needs to consider weakly symmetric ans\"atze, where time-reversal is broken (a PSG classification of such chiral ans\"atze together with a brief discussion can be found in Appendix \ref{appendix:WSansatz}). 
We did not perform an exhaustive search for bosonic chiral spin liquid states and leave this question open for future work.

Finally, increasing $J_K$ further, we observe another continuous transition from SL2 to the SL3 spin liquid phase. The minima of the spinon dispersion in SL3 are located at the M and $\Gamma$ points. We showed that we recover classical ground states of the Kitaev model on the triangular lattice in the large $S$ limit under reasonable assumptions, which further substantiates the validity of our SBMFT approach. The spin correlation tensor $S^{ab}(\vec{q})$ in SL3 is again diagonal, but the three diagonal components are not equal due to the lack of $SU(2)$ symmetry. 

A notable observation is that the gaps of all elementary excitations remain finite throughout the two continuous transitions between the three $Z_2$ spin liquid phases, whereas the gap of collective triplet or singlet excitations closes. This provides an interesting example of a continuous quantum phase transition between different symmetry enriched topological phases which do not differ in their projective symmetry group.

\acknowledgements
We thank M.~Garst,  J.C.~Halimeh, S.~Sachdev and M.~Vojta for helpful discussions.
P.K.~was supported by an Ad futura scholarship (No.~11010-234/2014) awarded by the Slovene Human Resources Development and Scholarship Fund. Moreover, we acknowledge support by the German Excellence Initiative via the Nanosystems Initiative Munich (NIM).


\appendix

\section{The form of the matrix $H_\vec{k}$}
\label{appendix:HMF}
Here we explicitly state the elements of the $4\times4$ matrix $H_\vec{k}$ in Eq.~\eqref{eq:hmf0}. Again we use the notation $k_1=\vec{k} \cdot \vec{a_x}$, $k_2=\vec{k} \cdot \vec{a_y}$ and $k_3=-k_1-k_2$.
{
\begin{align}
H_{ii} &= \frac{\lambda}{2} + \frac{(J_H + J_K)}{2} B (\cos k_1+\cos k_2 +\cos k_3 ) \nonumber\\
H_{14} &= \frac{1}{2} (i J_H A (\sin k_1 +\sin k_2 +\sin k_3 )- i J_K  t\cos k_3 ) \nonumber\\
H_{23} &= \frac{1}{2} (-i  J_H A(\sin k_1 +\sin k_2 +\sin k_3 ) - i J_K  t\cos k_3 ) \nonumber\\
H_{13} &= \frac{J_K}{2}  ( i \ t \cos k_1  - t\cos k_2 )\\
H_{24} &= \frac{J_K}{2}  (-i  \ t \cos k_1  - t\cos k_2 ) \nonumber
\end{align}
\begin{align}
H_{32} &= H_{23}^*  &   H_{41} &= H_{14}^* &
H_{31} &= H_{13}^*  &   H_{42} &= H_{24}^*  \nonumber\\
H_{12} &= 0 & H_{21}&= 0 & H_{34}&= 0 & H_{43}&= 0 \nonumber
\end{align}
}

\section{Weakly-symmetric ans\"atze}
\label{appendix:WSansatz}

We construct weakly-symmetric ans\"atze on the triangular lattice following the procedure introduced by Messio \emph{et al}.~[\onlinecite{PhysRevB.87.125127}] and extend it to include the triplet bond operators $\hat{t}^\gamma$ as defined in Eq.~\eqref{def of t}. The main difference is that the six-fold rotations around a triangular lattice site are replaced by pseudo-rotations in or case, where the spin operators transform as well under lattice rotations due to spin-orbit coupling. Details of the computation are published elsewhere.\cite{ourPSG}

A list of all possible weakly symmetric ans\"atze is shown in Table \ref{tab:summary psg}, where we use the same definitions and notation as in Ref.~[\onlinecite{PhysRevB.87.125127}]. Ans\"atze with $p_1=1$ have a doubled unit cell with extra $\pi$ phases on certain bonds, the integer $k$ refers to complex phases $\phi_A = 0,\ k \, 2 \pi/3,\ k \, 4\pi/3$ of the $A$ parameter on bonds in $\mathbf{a}_{x,y,z}$ direction, whereas $\phi_B$ and $\phi_t$ denote the phases of the $B$ and $t$ bond parameters (for $k\neq0$ the $t$ fields carry the phase $\phi_A+\phi_t$). The only difference in our notation is that we allow for negative moduli of the parameters, i.e.~phase differences of $\pi$ correspond to the same ansatz.

 In the main part of this paper we studied only the fully symmetric ansatz 1 from Tab.~\ref{tab:summary psg}.
Using the slightly generalized chiral ansatz 3 instead, which allows for complex $B$ parameters, doesn't lead to a different mean-field ground-state, because the self-consistency condition always yields $\phi_B=0$. 

A brief discussion of the properties of ansatz 4 follows below. We leave the detailed study of the more complicated ans\"atze 5 and 6 open for future work.

\begin{table}[h]
\centering
\begin{tabular}{|c|c|c|c|c|}
\hline
Ansatz	& $p_1$	&k	&$\phi_B$	&$\phi_t$\\
\hline
1	&0	&0	&0	&0	\\
\hline
2	&0	&0	&0 	&any 	\\
\hline
3	&0	&0	&any 	&0 	\\
4	&0	&-1,1	&0	&0\\
5	&1	&0	&$\pi/2$ 	&0\\
6	&1	&-1,1	&$\pi/2$ 	&0\\
\hline
\end{tabular}
\caption{Summary of all possible weakly symmetric PSG's. Here we allow for negative fields so phases 0 and $\pi$ correspond to the same ansatz.
Ansatz 1 respect all symmetries including time reversal and is the only totally symmetric ansatz, ansatz 2 breaks only time reversal symmetry.
Ans\"atze 2-6 are weakly symmetric and correspond to the chiral states, where parity symmetry is broken as well. Moreover, ans\"atze with $p_1=1$ have a doubled unit cell.
}
 \label{tab:summary psg}
\end{table}

\subsection*{Ansatz 4}
\label{appendix:Anstzes 4}

\begin{figure}[t]
\centering
 \includegraphics[width= \columnwidth]{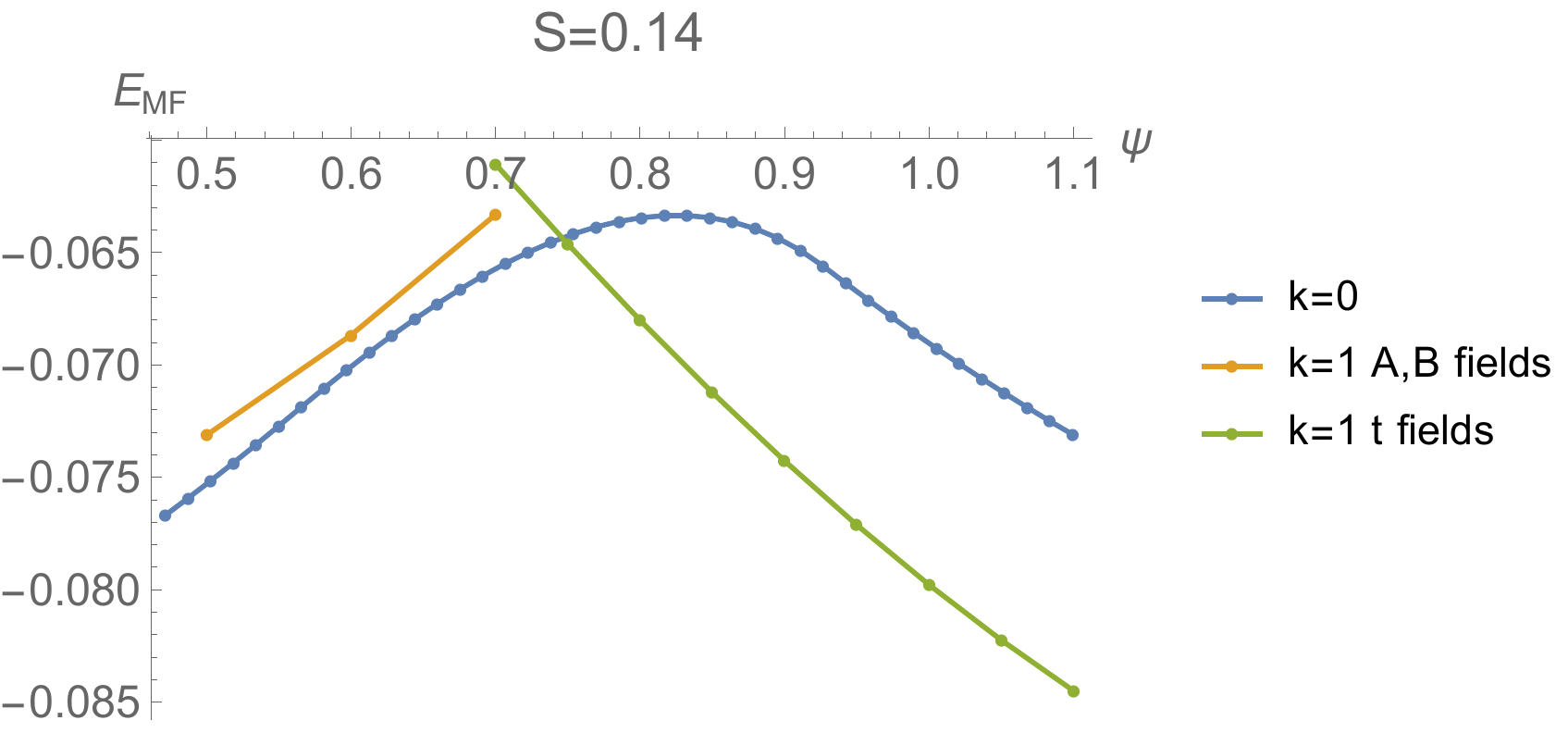}
\caption{(Color online) Mean-field ground state energy of ansatz 1 ($k=0$, discussed in the main text) vs.~the energy of the weakly symmetric ansatz 4 from Tab.~\ref{tab:summary psg} ($k=1$) for $S=0.14$. Ansatz 4 has a lower energy than ansatz 1 close to Kitaev point for $\psi \gtrsim 0.75$.
}
\label{fig:Emf k=1}
\end{figure}

Here we summarize relevant properties of ansatz No.~4 (characterized by $k=1$) as listed in Tab.~\ref{tab:summary psg}. 
Using the saddle point equations with ansatz 4, we find two different QSL phases as function of $\psi$, separated by a first order phase transition at $\psi \simeq 0.7$ for $S=0.14$. In the phase at small $\psi$ close to the Heisenberg point the $t$ fields vanish, whereas in the other phase around antiferromagnetic Kitaev point the $A$ and $B$ fields are zero at the saddle point. For $\psi\gtrsim0.75$ (at $S=0.14$) the mean-field ground state energy of ansatz 4 is actually lower than the energy of the fully symmetric ansatz 1, which we discussed in detail in the main text. A plot of the energies is shown in Fig.~\ref{fig:Emf k=1}.

\begin{figure}
\centering
 \includegraphics[width= 0.8 \columnwidth]{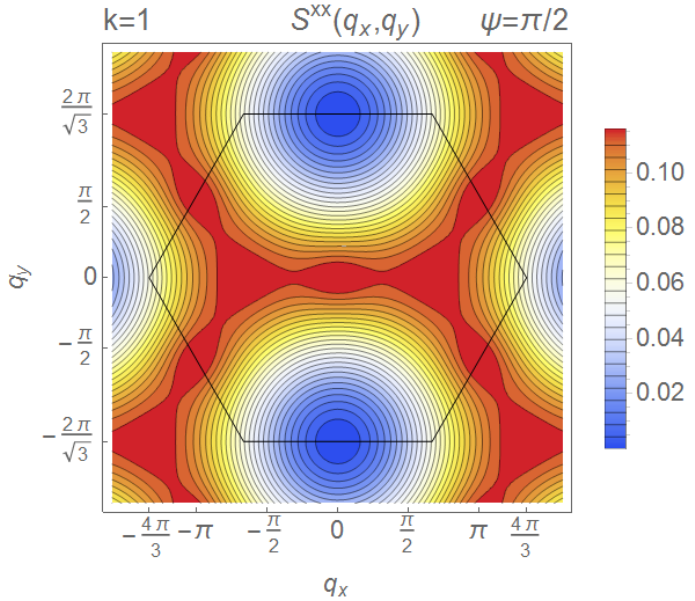}
\caption{(Color online) The xx-component of the static spin structure factor  $S^{xx}(\vec{q})$ for ansatz 4 at the Kitaev point. Note the similarities to $S^{xx}(\vec{q})$ in SL3 phase, shown in Fig.~\ref{fig:Sxx}.
}
\label{fig:k1Sxx}
\end{figure}

To characterize the chiral phase corresponding to ansatz 4 we calculated the xx-component of the spin structure factor $S^{xx}(\vec{q})$, shown in Fig.~\ref{fig:k1Sxx} and studied the magnetically ordered phase in the semi-classical limit at large $S$.

\begin{figure}
\centering
 \includegraphics[width= 0.8 \columnwidth]{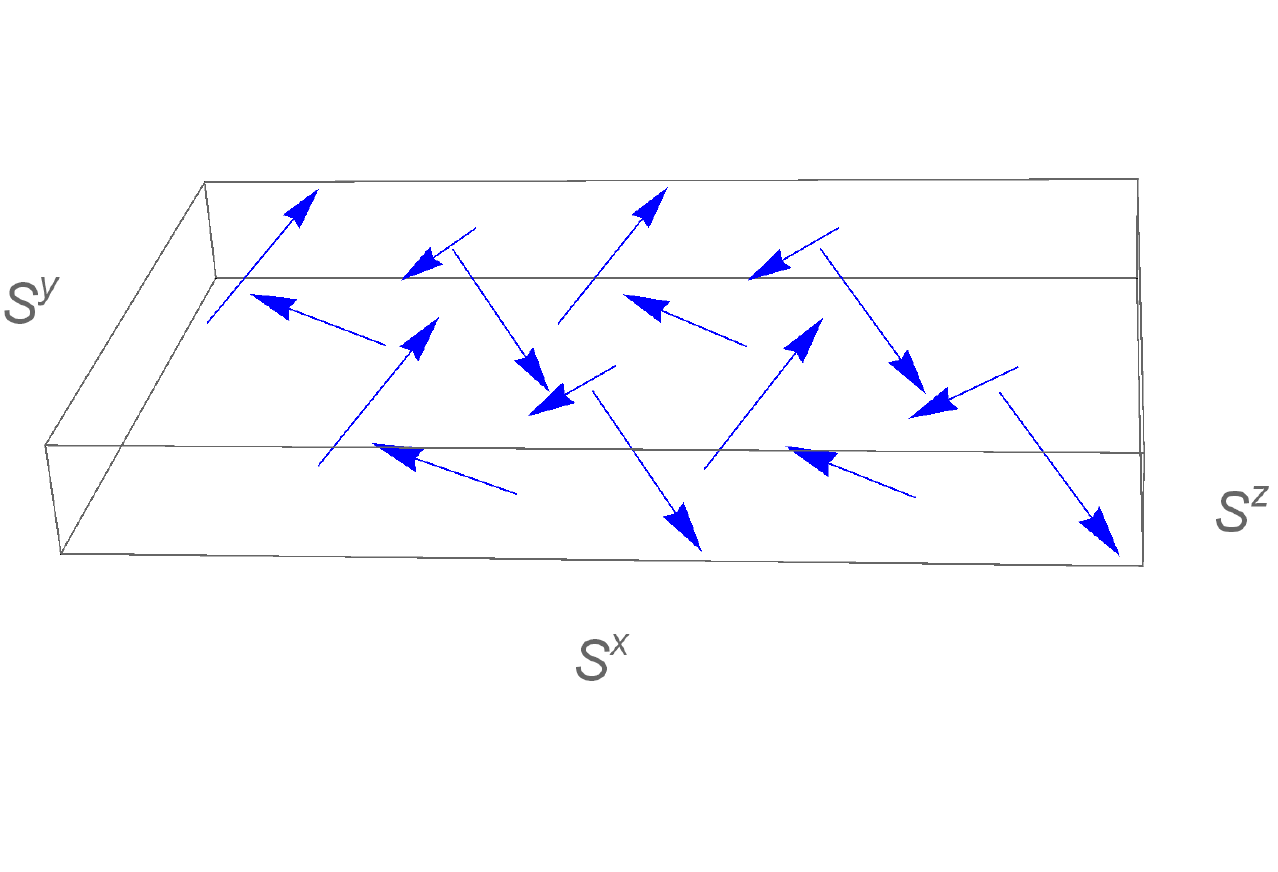}
\caption{(Color online) Ordered moment in the classical limit of ansatz 4 at the Kitaev point, which is indeed a non-coplanar classical ground state of the Kitaev model. 
}
\label{fig:k1classical}
\end{figure}

At the Kitaev point the minima of the spinon dispersion corresponding to ansatz 4 coincide with the dispersion minima of the SL3 state. Using the four eigenvectors of the Bogoliubov transformation matrix, consistency under exchanging momenta $\vec{k} \to \vec{-k}$ and further assuming a constant spin size and zero total magnetization, we obtain the following ordered moment in the classical limit:
\begin{align}
 \vec{S}(n, m)= \frac{1}{\sqrt{3}} ((-1)^{n+m}, -{(-1)}^{m}, (-1)^n).
\end{align}
where $n$ and $m$ again denote the position of the triangular lattice site $\vec{r}=n \vec{a_x} + m \vec{a_y}$.
A plot of this order parameter is shown in Fig.~\ref{fig:k1classical}. In contrast to the coplanar ordered moment obtained from ansatz 1, this spin configuration is non-coplanar, as expected from condensing spinons in a chiral spin liquid where time reversal symmetry is broken. Moreover, this spin configuration is also one of the degenerate classical ground states of the Kitaev model with energy $E=-J_\text{K} N$. Ansatz 4 thus recovers non-coplanar states in the degenerate ground-state manifold of the classical Kitaev model, which were missing in ansatz 1.

\bibliographystyle{apsrev4-1}
\bibliography{references}

\end{document}